\documentstyle[aps,tighten,graphicx,prl,twocolumn]{revtex}

\begin{document}
\title{Violation of multi-particle Bell inequalities for low and high flux
parametric amplification using both vacuum and entangled input states. \\
}
\author{M. D. Reid$^{1}$, W. J. Munro$^{2}$ and F. De Martini$^{3}$}
\address{1) Physics Department, University of Queensland, Brisbane, Australia,\\
2) Hewlett-Packard Laboratories, Filton Road, Stoke Gifford, Bristol, 
BS34 SQ2, United Kingdom, \ \ 
\\
3) Dipartimento di Fisica and INFM, Universit\`{a} di Roma ''La Sapienza'',\\
Roma 00185, Italy.\\
}
\date{\today}
\maketitle

\begin{abstract}
We show how polarisation measurements on the  
output fields generated by parametric down conversion
 will reveal  a violation of 
multi-particle Bell inequalities, in the regime of both low and high output
intensity.  In this case each spatially separated system, upon which a 
measurement is performed, is comprised 
of more than one particle. In view of the formal analogy with spin 
systems, the proposal provides an opportunity to 
test the predictions of quantum mechanics for spatially separated 
higher spin states. Here the  
quantum behaviour possible even where 
measurements are performed on systems of large quantum (particle) 
number may be demonstrated. 
 Our proposal applies to both vacuum-state signal and idler
inputs, and also to the quantum-injected parametric amplifier as studied by De
Martini et al. The effect of detector inefficiencies is included.
\end{abstract}

\vskip 1 truecm

\vskip 1 truecm

\narrowtext
\vskip 0.5 truecm

\section{Introduction}

There is increasing evidence for the failure of ``local realism'' as defined
originally by Einstein, Podolsky and Rosen$^{\cite{EPR}}$, Bohm$^{\cite{Bohm}
}$ and Bell$^{\cite{Bell,CS,GHZ}}$. For certain correlated quantum systems,
Einstein, Podolsky and Rosen (EPR) argued in their famous 1935 EPR paradox 
that ``local realism'' is sufficient to
imply that the results of measurements are predetermined. These
predetermined ``hidden variables'' exist to describe the value of a physical
variable, whether or not the measurement is performed, and as such are not
part of a quantum description. Bell later showed that the predictions of
quantum mechanics for certain ideal quantum states could not be compatible
with such local hidden variable theories. It is now widely accepted 
therefore, as
a result of Bell's theorem and related experiments$^{\cite{exp}}$, that
local realism must be rejected.

Recently three-photon states demonstrating a contradiction of 
quantum mechanics with local hidden variables have been 
generated$^{\cite{3photon}}$. 
 A multi-particle entanglement involving four trapped ions 
 has also 
 been recently 
realized by Sackett et al $^{\cite{wine4}}$,  
and for atoms and photons in  
cavities by Rauschenbeutel et al$^{\cite{Haroche}}$. These 
experiments involve measurements performed on separated 
subsystems that are microscopic. 
Recently the EPR paradox, itself a demonstration of entanglement,
 has been realized where each measurement is performed on a 
 macroscopic system. Such  
experiments were performed initially by Ou et al$^{\cite{eprexpou}}$ 
using intracavity 
parametric oscillation below threshold, and have now been achieved 
for intense fields using parametric oscillation 
above threshold by Zhang et al$^{\cite{eprexpzhang}}$, 
and for pulsed fields by Silberhorn et al$^{\cite{eprexpsil}}$. 
There have been further theoretical proposals to demonstrate the 
macroscopic nature of  EPR correlations$^{\cite{eprmr,eprtomb}}$.
However to our knowledge,
 experimental efforts using clearly spatially separated systems,  
 testing local realism directly through a 
violation of a Bell-type inequality, (or through the  
Greenberger-Horne-Zeilinger effect$^{\cite{GHZ}}$), have so far been confined
to the most microscopic of systems, where each measurement is made on 
a system comprising only one particle.

A theoretical demonstration of a predicted incompatibility of quantum
mechanics with local hidden variable theories
 for systems of potentially more than one
particle per detector came with the work of Mermin$^{\cite{Mermin}}$, 
Mermin and Garg$^{\cite{Mermin}}$ and  
Mermin and Schwarz$^{\cite{MerminSchwarz}}$ who
showed violations of Bell inequalities to be possible 
for a pair of spatially separated higher spin $j$ particles, where 
$j$ can be arbitrarily large. The violation of a Bell inequality for
multi-photon macroscopic systems was put forward by 
Drummond$^{\cite{Drum}}$.  Such manifestations of irrefutably 
quantum behaviour are contradictory to the notion that classical 
behaviour is obtained in the limit where the quantum numbers, or 
particle numbers, become large. The work of Peres$^{\cite{Peres}}$ 
has shown how 
the transition to classical behaviour (local 
realism) is obtained through measurements that become 
increasingly fuzzy. 
To observe the failure of local 
realism it is generally necessary to perform measurements 
sufficiently accuracy so as to resolve the $2j+1$ eigenvalues. 
The contradiction of quantum mechanics with local realism for
multi-particle or higher spin systems has since been explored 
theoretically in a number of
works$^{\cite{multitheo,BW,MB}}$.

In this paper we present a proposal to test for multi-photon violations of
local realism, by way of a violation of a Bell inequality, using parametric
down conversion. 
 Our proposal involves a four-mode parametric 
interaction, considered initially by Reid and Walls$^{\cite{4parexppro}}$ 
and Horne et al $^{\cite{4parexppro}}$, 
as may be generated for example using two parametric  
amplifiers, or using two competing parametric processes. 
Such parametric interactions were used 
to demonstrate  experimentally violations of a Bell-type inequality
 for the single photon 
case by Rarity and Tapster$^{\cite{4parexppro}}$, and there has been 
further experimental work $^{\cite{4parexppro,exp}}$.   
While initially we consider vacuum inputs with two parametric amplifiers, 
our proposal is 
also  
formulated for the specific configuration of the quantum injected parametric
amplifier$^{\cite{deM}}$. Here `` multi-particle Bell 
inequalities'' refer to  
Bell-inequality tests applying to situations where each 
measurement is performed on a system of more
than one particle. In our proposal the measurement is of the number of 
particles polarised ``up'' minus the number of particles polarised 
``down''.  Because of the formal analogy to 
a pair of spin $j$ particles, our proposal allows a test of 
the predictions of quantum mechanics for the higher spin states.  

We will focus on two regimes of experimental operation. The first
corresponds to relatively low interaction strength so that the mean
signal/idler output is small and we have low incident photon numbers on
polarisers which serve as the measurement apparatus. 
Here it is shown how certain measured probabilities of detection
of precisely $n$ photons transmitted through the polariser can violate local
realism, and represent a test of the established higher-spin 
 results.
 Previous calculations$^{\cite{MB}}$ of this 
type were primarily confined to situations
of extremely low detection efficiency. Here the results are presented for
higher efficiencies more compatible with current experimental proposals.

Our second regime of interest is that of higher output signal/idler intensity,
where many photons fall incident on the measurement apparatus. We
present a proposal for a violation of a Bell inequality, where one measures
the probability of a range of intensity output through the polariser. The
application of Bell inequality theorems, and the effect of detection
inefficiencies on the violations predicted, to situations where many photons
fall on a detector is relevant to the question of whether or not tests of
local realism can be conducted in the experiments such as those performed by
Smithey et al$^{\cite{Smithey}}$. In the Smithey et al experiment,
 correlation of photon number between
two spatially separated but very intense fields is sufficient to give
``squeezed'' noise levels. Previous studies by Banaszek and 
Wodkiewicz $^{\cite{BW}}$ have demonstrated violations of Bell inequalities to be 
possible for certain measurements for the 
signal/idler outputs of the parametric amplifier. In these high flux experiments detection losses
can be relatively small on a percentage basis, as compared to traditional
Bell inequality experiments involving photon counting with low incident
photon numbers. The exact sensitivity of the violations to loss determines
the feasibility of a multi-particle, no-loop-hole violation of a Bell inequality.

\section{Derivation of the multi-particle Bell inequalities}

We consider a general situation as depicted in Figure 1 of two pairs of
spatially separated fields. The two modes at location $A$ are denoted by the
boson operators $a_{1}$ and $b_{1}$, while the two modes at location $B$,
spatially separated from $A$, are denoted by the boson operators $a_{2}$ and 
$b_{2}$. One can measure at $A$ the photon numbers $c_+^\dagger c_+$ and $%
c_-^\dagger c_-$; and similarly at $B$ one can measure, simultaneously, the
photon numbers $d_+^\dagger d_+$ and $d_-^\dagger d_-$, where 
\begin{eqnarray}
c_{+} &=&a_{1}\cos \theta +b_{1}\sin \theta  \nonumber \\
c_{-} &=&-a_{1}\sin \theta +b_{1}\cos \theta  \nonumber \\
d_{+} &=&a_{2}\cos \phi +b_{2}\sin \phi  \nonumber \\
d_{-} &=&-a_{2}\sin \phi +b_{2}\cos \phi  \label{eqn:modes}
\end{eqnarray}
These measurements may be made$^{\cite{exp,4parexppro}}$ with the use of two
sets of polarisers, to produce the transformed fields $c_+$ and $d_+$,
followed by photodetectors at $A$ and $B$ to determine the photon numbers $%
c_+^\dagger c_+$ and $d_+^\dagger d_+$ respectively. We note that each
measurement at $A$ corresponds to a certain choice of parameter $\theta$.
Similarly a measurement at $B$ corresponds to a certain choice of $\phi$. 
In our final proposal, the fields $a_{1}$ and 
$a_{2}$ will be the correlated signal/idler 
outputs of a single parametric amplifier with 
Hamiltonian $H=i\hbar g (a_1^{\dagger }a_2^{\dagger } - 
 a_1a_2)
$, while $b_{1}$ and 
$b_{2}$ are the outputs of a second parametric amplifier with 
Hamiltonian $H=i\hbar g (b_1^{\dagger }b_2^{\dagger }
 -  b_1b_2)$.

\begin{figure}
  \includegraphics[scale=.45]{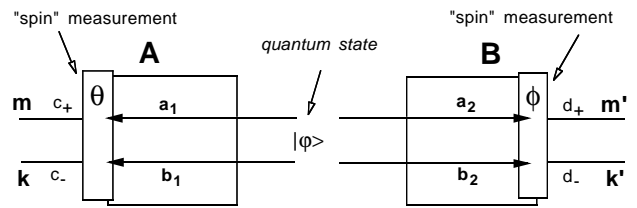}\\
\caption{Schematic diagram of the experimental arrangement to test the Bell
inequality. Here $m,k,m^{\prime}k^{\prime}$ are the results of measurement
of $c_+^\dagger c_+$, $c_-^\dagger c_-$, $d_+^\dagger d_+$, $d_-^\dagger d_-$
respectively. Binary outcomes $+1$ and $-1$ are defined and we measure joint
and marginal probabilities $P_{++}^{AB}(\protect\theta,\protect\phi)$, $%
P_{+}^{A}(\protect\theta)$ and $P_{+}^{B}(\protect\phi)$ for obtaining $+1$.}
\label{fig1}
\end{figure}

Let us denote the outcome of the photon number measurements $c_+^\dagger c_+$%
, $c_-^\dagger c_-$, $d_+^\dagger d_+$, and $d_-^\dagger d_-$ as $m$, $k$, $%
m^{\prime}$ and $k^{\prime}$ respectively. We will classify the result of
our measurements made at each of $A$ and $B$ as one of two possible
outcomes. For certain outcomes $m,k$ at $A$ we will assign the value $+1$.
(This choice of outcomes will be specified later). Otherwise our result is $%
-1$. Similarly at $B$, certain values $m^{\prime},k^{\prime}$ are classified
as result $+1$, while all other outcomes are designated $-1$. This binary
classification of the results of the measurement is chosen to allow an easy
application of Bell's theorem.

To establish Bell's result, one considers joint measurements where the
photon numbers $c_+^\dagger c_+$, $c_-^\dagger c_-$ and $d_+^\dagger d_+$, $%
d_-^\dagger d_-$ are measured simultaneously at the spatially separated
locations $A$ and $B$ respectively. A joint measurement will give one of
four outcomes, $+1$ or $-1$ for each particle. By performing many such
measurements over an ensemble, one can experimentally determine the
following: $P_{++}^{AB}\left( \theta ,\phi \right) $ the probability of
obtaining $+1$ for particle $A$ and $+1$ for particle $B$ upon simultaneous
measurement with $\theta$ at $A$ and $\phi $ at $B$; $P_{+}^A\left( \theta
\right)$ the marginal probability for obtaining the result $+1$ upon
measurement with $\theta$ at $A$; and $P_{+}^B\left( \phi \right)$ the
marginal probability of obtaining the result $+1$ upon measurement with $%
\phi $ at $B$.

Assuming a general local hidden variable theory then, we can write the
measured probabilities as follows. 
\begin{eqnarray}
P_{+}^A(\theta )=\int \rho(\lambda) \quad p_{+}^A(\theta, \lambda ) \quad
d\lambda
\end{eqnarray}
The probability of obtaining `+1' for $S_\phi ^B$ is 
\begin{eqnarray}
P_{+}^B(\phi )=\int \rho(\lambda)\quad p_{+}^B(\phi, \lambda ) \quad d\lambda
\end{eqnarray}
The joint probability for obtaining `+1' for both of two simultaneous
measurements with $\theta$ at $A$ and $\phi$ at $B$ is 
\begin{eqnarray}
P_{++}^{AB}(\theta ,\phi )= \int \rho(\lambda) \quad p_{+}^A(\theta, \lambda
) p_{+}^B(\phi, \lambda )\quad d\lambda
\end{eqnarray}
Here $\theta $ and $\phi $ denote the choice of measurement at the locations 
$A$ and $B$ respectively. The independence of $p_{+}^A(\theta, \lambda )$ on 
$\phi $, and $p_{+}^B(\phi, \lambda )$ on $\theta$, follows from the
locality assumption. The measurement made at $B$ cannot instantaneously
influence the system at $A$.

It is well known$^{\cite{Bell,CS}}$ that one can derive the following
``strong'' Bell-Clauser-Horne-Shimony-Holt inequality from the assumptions of local
realism made so far. 
\begin{eqnarray}
S&=&{\frac{{P_{++}^{AB}(\theta,\phi)-P_{++}^{AB}(\theta,\phi^{%
\prime})+P_{++}^{AB}(\theta^{\prime},\phi)
+P_{++}^{AB}(\theta^{\prime},\phi^{\prime})}}{{P_{+}^{A}(\theta^{%
\prime})+P_{+}^{B}(\phi)}}} \nonumber\\
 &\leq&1  \label{eqn:Bellineq}
\end{eqnarray}

\section{Multi-particle ``spin'' state violating the Bell inequalities}

Bell inequality violations have been proposed previously for macroscopic or
multi-particle states$^{\cite{Mermin,MerminSchwarz,Drum,Peres,multitheo,MB}}$. 
Previous studies by Mermin, Peres and others have considered 
violations by states
of arbitrary spin $j$. There is a formal equivalence by way of the 
Schwinger representation to bosonic states of $N=2j$ photons$%
^{\cite{MB}}$. For example we consider the following $N$ particle state 
\begin{eqnarray}
|\varphi_{N}\rangle = {\frac{{1}}{{\ N! \left(N+1\right)^{1/2}}}}
\left(a_1^\dagger a_2^\dagger+b_1^\dagger b_2^\dagger\right)^N |0\rangle
|0\rangle  \label{eqn:state}
\end{eqnarray}
where the boson operators $a_{1}$ and $b_{1}$ are as in section 2 and figure
1. This state was presented, and shown to violate local realism where 
each measurement is performed on systems of   
$N$ particles (where $N$ can be macroscopic), by Drummond$^{\cite{Drum}}$. 
We introduce the Schwinger spin operators 
\begin{eqnarray}
S_x^A &=&(a_1b_{1}¥^{\dagger }+a_1^{\dagger }b_{1}¥)/2  \nonumber \\
S_y^A &=&\left( a_1b_{1}¥^{\dagger }-a_1^{\dagger }b_{1}¥\right) /2i \nonumber\\
S_z^A&=&(b_{1}¥^{\dagger }b_{1}¥-a_1^{\dagger }a_1)/2 \nonumber\\
S_x^B &=&(a_{2}¥b_2^{\dagger }+a_{2}¥^{\dagger }b_2)/2  \nonumber \\
S_y^B &=&\left( a_{2}¥b_2^{\dagger }-a_{2}¥^{\dagger }b_2\right) /2i 
\nonumber\\
S_z^B&=&(b_2^{\dagger }b_2-a_{2}¥^{\dagger }a_{2}¥)/2 \label{eqn:sch}
\end{eqnarray}
The
photon number difference measurements at each detector corresponds in this
formalism to a measurement of the ``spin'' component 
\begin{eqnarray}
S_z^A\left( 2\theta \right) &=&\left( c_{+}^{\dagger }c_{+}-c_{-}^{\dagger
}c_{-}\right) /2  \nonumber \\
S_z^B\left( 2\phi \right) &=&\left( d_{+}^{\dagger }d_{+}-d_{-}^{\dagger
}d_{-}\right) /2  \label{eqn:rot}
\end{eqnarray}
as determined by the polariser angle $\theta$ or $\phi$. Here 
$S_{z}^{A}¥(2\theta)=S_{z}^{A}\cos 2\theta + 
S_{x}^{A}¥\sin 2\theta$ and $S_{z}^{B}¥(2\phi)=S_{z}^{B}\cos 2\phi + 
S_{x}^{B}¥\sin 2\phi$. 
The quantum state (\ref{eqn:state}) can be written as 
\begin{eqnarray}
|\varphi_{N}\rangle = {\frac{{1}}{{ \left(2j+1\right)^{1/2}}}}
\sum_{m=-j}^{+j}|j,m\rangle_{A}|j,m\rangle_{B}  
\end{eqnarray}
where $|j,m\rangle_{A}$ and $|j,m\rangle_{B}$ are the usual eigenstates of 
$S^{2}_{A}$,$S_{z}^{A}$, and $S^{2}_{B}¥$,$¥S_{z}^{B}$ respectively,
 and $j=N/2$. The singlet state 
 \begin{eqnarray}
|\varphi_{N}\rangle = {\frac{{1}}{{ \left(2j+1\right)^{1/2}}}}
\sum_{m=-j}^{+j}(-1)^{j-m}|j,m\rangle_{A}|j,-m\rangle_{B}  
\label{eqn:sing}
\end{eqnarray}
 studied by previous authors 
 is obtained upon substituting 
 $a_{1}$ with $-a_{1}$, and interchanging $a_{2}$ and $b_{2}$ in the 
 definitions of $S_{x}^{B}, S_{y}^{B}$ and $S_{z}^{B}$. 
 The predictions as given in this paper of the quantum state  (\ref{eqn:state}) with 
 measurements  (\ref{eqn:sch}) and  (\ref{eqn:rot}) using particular $\theta$ and $\phi$ 
 will be identical to 
 the predictions of the singlet state  (\ref{eqn:sing}) 
 above with measurements (\ref{eqn:sch}) and  (\ref{eqn:rot})
  but replacing $\phi$ and 
 $\theta$ with $\phi_{spin}¥$ and 
 $\theta_{spin}$ where  
 $2\phi_{spin}=2\phi+\pi$ and $\theta_{spin}=-\theta$.  
\begin{figure}
  \includegraphics[scale=.45]{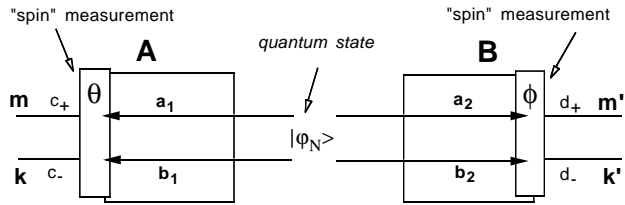}\\
\caption{Experimental arrangement to test the Bell inequality for state (6),
in the absence of loss. Our outcome at $A$ is designated $+1$ if $m\geq fN$,
and $+1$ for $B$ if $m^{\prime}\geq fN$ , where $f$ is a preselected
fraction. Predicted violations of the Bell inequality (5) are shown in
Figure 3.}
\label{fig2}
\end{figure}

For the purpose of our particular experimental proposal we first demonstrate
the failure of multi-particle local realism for the $N$-states (\ref
{eqn:state}) as follows. We choose the following binary classification of
outcomes. If the result $m$ of the photon number measurement $c_+^\dagger
c_+ $ is greater than or equal to a certain fraction $f$ of the total photon
number $m+k$ detected at $A$, then we have the result $+1$. Otherwise our
result is $-1$. The outcome of a measurement at the location $B$ is
classified as $+1$ or $-1$ in a similar manner.

A perfect correlation between $c_+^\dagger c_+$ and $d_+^\dagger d_+$ is
predicted for the state (\ref{eqn:state}) for $\theta=\phi$, a result $n$
for $c_+^\dagger c_+$ implying a result $n$ for $d_+^\dagger d_+$. For such
situations of perfect correlation, we are able to deduce, if we assume local
realism, following the reasoning of Einstein, Podolsky and Rosen$^{\cite{EPR}%
}$, the existence of a set of ``elements of reality'', $m_\theta ^A$ and $%
m_\phi ^B$, one for each subsystem at $A$ and $B$, and one for each choice
of measurement angle, $\theta$ or $\phi$ at $A$ or $B$ respectively. The
whole set of ``elements of reality'' ${m_{\theta}^A}$ and ${m_{\phi}^B}$
form a set of ``hidden variables'' representing the predetermined value of
the results of ``spin'' measurements which can be attributed to the two
particle system at a given time. This local realism assumption then implies
the inequality (\ref{eqn:Bellineq}).

The averages (\ref{eqn:Bellineq}) can be calculated using quantum mechanics
for the state (\ref{eqn:state}) and the quantity $S$ computed. Violations of
the Bell inequality (\ref{eqn:Bellineq}) are found for a range of parameters
as illustrated in Figure 3. Here we have selected the following relation
between the angles: $\phi -\theta=\theta^{\prime}-
\phi=\phi^{\prime}-\theta^{\prime}=\psi$ and $\phi^{\prime}-\theta=3\psi$.
This combination has been shown to be optimal for the cases $N=1$$^{\cite
{Bell,CS}}$ and for all $N$ values with $f=1$$^{\cite{Drum}}$. For each $f$,
the value of $S$ tends to an asymptotic value as $N$ increases.

The violation of the Bell inequality (\ref{eqn:Bellineq}) is greatest for $%
f=1$, where our result $+1$ at $A$, for example, corresponds to detecting
all $N$ photons in the $c_+$ mode. This result has been presented previously$%
^{\cite{Drum}}$. While this value of $f$ gives the strongest violation, the
actual probability of the $+1$ event in this case becomes increasingly small
as $N$ increases especially if detection inefficiencies are to be included
as in later calculations. From this point of view, to look for the most
feasible macroscopic experiment, the violations with reduced $f$ become
important. We see that the magnitude of violation decreases with increasing $%
f$, so that the asymptotic value at $f=.5$ is 1.

\begin{figure}
   \includegraphics[scale=.7]{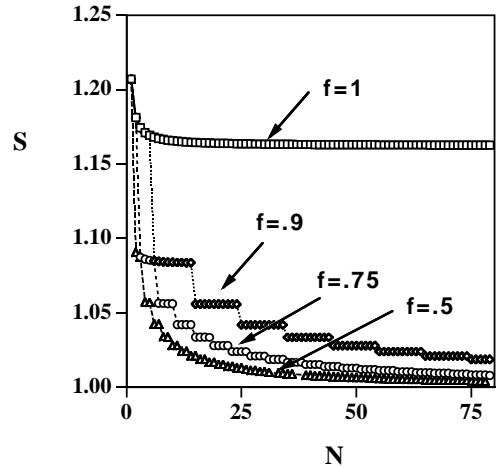}\\
\caption{Plot of $S$ showing violation of the Bell inequality (5) (and (16))
versus $N$ for the quantum state (6). The results are optimised with respect
to the angle $\protect\psi$ as defined in the text. A violation is obtained
when $S>1$. For $f=1$, the optimal angle $\protect\psi$ is $%
.39,3.4,.22,.19,3.1$ for $N=1,2,3,4,80$ respectively. Results for values of $%
f=0.5-x$ are identical to those for $f=0.5+x$. }
\end{figure}

\section{The effect of detection inefficiencies: derivation of a weaker Bell
inequality}

The effect of loss through detection inefficiency is important, since this
limits the experimental feasibility of a test of the Bell inequality. To
date to our knowledge 
the ``strong'' inequality of the type (\ref{eqn:Bellineq}) has not yet been
violated$^{\cite{CS}}$ in any experiment involving photodetection,
because of the detection inefficiencies which occur in photon counting
experiments, although recent experiments by Rowe et al 
$^{\cite{exp}}$ violate a true  
Bell inequality 
for trapped ions, with limited spatial separation.  

It is well documented$^{\cite{Bell,CS}}$ that it is possible to derive, with
the assumption of additional premises, a weaker form of the Bell-
Clauser-Horne inequalities which have been violated in single photon
counting experiments. Before proceeding to derive a ``weak'' Bell inequality for
multi-particle detection, we outline the effect of detection inefficiencies
on the violation, as shown in Figure 2, of the strong Bell inequality (\ref
{eqn:Bellineq}).

We introduce a transmission parameter $T$, defining $T$ as the probability that a single
incoming photon will be detected, the intensity of the incoming field being
reduced by the factor $T$. $T$ is directly related to the detector
efficiency $\eta$ according to $T=\eta ^2$. We model loss in the standard
way by considering the measured field to be the transmitted output of a 
imaginary beam
splitter with the input being the actual quantum field incident on the
detector. The second input to the imaginary beam splitter is a vacuum field.
Calculating the probabilities of this measured field is equivalent to using
standard photocounting formulae which incorporate detection inefficiencies.

The following expression gives the final measured probability $%
P(m,k,m^{\prime},k^{\prime})$ for obtaining results $m,k,m^{\prime},k^{%
\prime}$ upon measurement of $c_+^\dagger c_+$, $c_-^\dagger c_-$ and $%
d_+^\dagger d_+$, $d_-^\dagger d_-$ respectively. Here $%
P_{Q}(m_{0},k_{0},m^{\prime}_{0},k^{\prime}_{0})$ is the quantum probability
for obtaining $m_{0},k_{0},m^{\prime}_{0},k^{\prime}_{0}$ photons, upon
measurement of $c_+^\dagger c_+$, $c_-^\dagger c_-$ and $d_+^\dagger d_+$, $%
d_-^\dagger d_-$, in the absence of detection losses. This quantum
probability is derivable from (\ref{eqn:state}). 
\begin{eqnarray}
P(m,k,m^{\prime},k^{\prime})&=&T^{m+k+m^{\prime}+k^{\prime}}%
\sum_{r,q,s,t=0}^{\infty} (1-T)^{r+q+s+t}  \nonumber \\
&\times&C_{r}^{m+r} C_{q}^{k+q} C_{s}^{m^{\prime}+s} C_{t}^{k^{\prime}+t} 
\nonumber \\
&\times& P_{Q}(m+r,k+q,m^{\prime}+s,k^{\prime}+t)  \label{eqn:loss}
\end{eqnarray}
Here $C_{r}^{m+r}=(m+r)!/r!m!$, and $r,q,s,t$ represent the number of
photons lost. We also consider the measured marginal probability. 
\begin{eqnarray}
P^{A}(m,k)&=&T^{m+k}\sum_{r,q=0}^{\infty}(1-T)^{r+q} C_{r}^{m+r} C_{q}^{k+q}
\nonumber \\
&\times& P_{Q}^{A}(m+r,k+q)  \label{eqn:margloss}
\end{eqnarray}
where $P_{Q}^{A}(m+r,k+q)$ represents the quantum probability for obtaining $%
m_{0},k_{0}$ photons upon measurement of $c_+^\dagger c_+$, $c_-^\dagger c_- 
$ in the absence of detection losses. This marginal quantum probability is
derivable from (\ref{eqn:state}).

With loss present there is a distinction between our actual quantum photon
number $m_{0}$ present on the detectors, and the final readout photon number $m$, which is taken to
be the result of the photon number measurement. (We must have $m\leq m_{0}$%
). Therefore a number of quantum probabilities will contribute in the 
calculation for the final
measured probability. This complicating 
effect may be avoided in the following manner.
The outcome at $A$ is labeled $+1$ only if $m\geq fN$ and $m+k=N$; and at $%
B $ if $m^{\prime}\geq fN$ and $m^{\prime}+k^{\prime}=N$. For $N$ photons
detected at each location $A$ or $B$, we are restricted to the outcomes
satisfying $m+k=m^{\prime}+k^{\prime}=N$ where loss has not occurred, for
the given initial quantum state $|\varphi _{N} \rangle $. In this situation
we get for the measured probabilities (\ref{eqn:loss}) 
\begin{eqnarray}
&\quad&P(m,N-m,m^{\prime},N-m^{\prime})=  \nonumber \\
&\quad&T^{2N} P_{Q}(m,N-m,m^{\prime},N-m^{\prime})  \label{eqn:lossamend}
\end{eqnarray}
and for the marginal 
\begin{eqnarray}
P^{A}(m,N-m)&=& T^{N} 
 P_{Q}^{A}(m,N-m)  \label{eqn:lossamendmarg}
\end{eqnarray}
Here $P_{Q}(m,N-m,m^{\prime},N-m^{\prime})$ is
the quantum probability (in the absence of loss) that measurement of $%
c_{+}^{\dagger }c_{+}$ and $d_{+}^{\dagger }d_{+}$, for the state $|\varphi
_{N} \rangle $ of equation (\ref{eqn:state}), will give results 
$m$ and $m^{\prime}$ respectively. 
This quantum probability is calculated from the quantum amplitudes 
$C_{m,m^{\prime}}^{(N)}=
 \langle \varphi_{N}|m\rangle _{\theta}¥|m^{\prime}\rangle_{\phi}$,
where $|m\rangle_{\theta}$, $|m^{\prime}\rangle_{\phi}$ are eigenstates of 
$c_+^\dagger c_+$, $d_+^\dagger d_+$ respectively,
and is given by 
$P_{Q}(m,N-m,m^{\prime},N-m^{\prime})=|C_{m,m^{\prime}}^{(N)}|^2$
The quantum marginal
for $|\varphi _{N} \rangle $ is $P_{Q}^{A}(m,N-m)=\sum_{m^{%
\prime}=0}^{N}|C_{m,m^{\prime}}^{(N)}|^2$.

The crucial effect of detection losses is that each measured joint
probability contains the factor $T^{2N}$ where $2N$ is the total number of
photons $m+k+m^{\prime}+k^{\prime}$ detected. This implies immediately
extreme sensitivity of the multi-particle strong Bell inequality (\ref
{eqn:Bellineq}) to loss, since this inequality involves the marginal which
scales as $T^{N}$. In the presence of loss $T$, the
 new predicted value for $S$  
 (required to test the strong Bell inequality (\ref{eqn:Bellineq})) 
is $T^{N}S_{0}¥$ where $S_{0}$ is 
the value ``S'' for $\varphi_{N}$ in the absence of loss as
 given graphically in Figure 3.
 It is seen then that we require $T$ to be $\sim
(1/S_{0})^{1/N}$ or larger in order to obtain 
the violations of the no loop-hole
inequalities (\ref{eqn:Bellineq}). For $N=2$ $S_{0}=1.18$, and 
this requires at least $T>\sqrt{%
1/1.18}=.92$. This figure is at the limits of current
technology, and compares with the requirement $T>.83$ for $N=1$.

\begin{figure}
  \includegraphics[scale=.45]{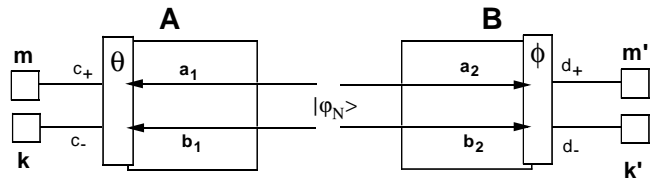}\\
\caption{Schematic diagram of the experimental arrangement to test the Bell
inequality for $|\protect\varphi_{N}\rangle$ with detection inefficiencies
present. Our outcome is $+1$ at $A$ if $m\geq fN$ and $m+k=N$; and $+1$ at $%
B $ if $m^{\prime}\geq fN$ and $m^{\prime}+k^{\prime}=N$. A violation of the
no auxiliary-assumptions Bell inequality (5) is possible only for high 
detector efficiencies $T=%
\protect\eta^{2}$. Predicted violations of the weak inequality (16) are as
for Figure 3.}
\label{fig4}
\end{figure}

We now derive a multiparticle form of the weaker inequality so that we can
also examine situations of significant detection loss. The result at $A$ is $%
+1$ if the number of photons $m$ detected at $c_{+}$ is $fN$ or more, and if
the total number of photons $m+k$ detected at $A$ satisfies $m+k=N$; $%
p_{+}^{A}(\theta,\lambda)$ is the probability of this event given the hidden
variable description $\lambda$. We define a new probability, $p_{+}^{A}(-,
\lambda)$, that the total photon number $m+k$ (at location $A$) is $N$,
given that the system is described by the hidden variables $\lambda$. This
total probability is then assumed to be independent of the choice of
polariser angle $\theta$ at $A$. Similarly we define a $p_{+}^{B}(-,
\lambda) $, the probability that the total number of photons $%
m^{\prime}+k^{\prime}$ at $B$ is $N$. This total probability is then assumed
to be independent of the polariser angle $\phi$ at $B$. We postulate as an
additional premise that the hidden variable theories will satisfy 
\begin{eqnarray}
p_{+}^{A}(\theta, \lambda) \leq p_+^A (-, \lambda)  \nonumber \\
p_{+}^{B}(\phi, \lambda) \leq p_+^B (-, \lambda). 
\label{eqn:weakassump}
\end{eqnarray}
Using the procedure and theorems of the previous works of Clauser and Horne$%
^{\cite{CS}}$ one may derive from the postulate of local realism and
assumption (\ref{eqn:weakassump}) the following ``weak'' Clauser-Horne-Shimony-Holt Bell
inequality, where the marginals are replaced by ``one-sided'' 
joint probabilities. Violation of this ``weaker'' Bell-CHSH inequality  
will only eliminate local hidden 
variable theories satisfying the auxiliary assumption. 
(\ref{eqn:weakassump}). 
\begin{eqnarray}
&\quad&S_{W}=\nonumber\\
&\quad&{\frac{{P_{++}^{AB}(\theta,\phi)-P_{++}^{AB}(\theta,\phi^{%
\prime})+P_{++}^{AB}(\theta^{\prime},\phi)
+P_{++}^{AB}(\theta^{\prime},\phi^{\prime})}}{{P_{++}^{AB}(\theta^{%
\prime},-)+P_{++}^{AB}(-,\phi)}}}\nonumber\\
&\quad &\leq 1  \label{eqn:weakineq}
\end{eqnarray}
Here we have defined new ``one-sided'' experimental joint probabilities 
as follows:
 $P_{++}^{AB}(\theta^{\prime},-)$, the joint probability of obtaining $%
+1 $ at $A$, with the polariser at $A$ set at $\theta^{\prime}$, and of
obtaining a total of $m^{\prime}+k^{\prime}=N$ photons at $B$. The joint
probability $P_{++}^{AB}(-,\phi)$ is the probability of obtaining a total of 
$m+k=N$ photons at $A$, and of obtaining $+1$ at $B$, with the polariser at $%
B$ set at $\phi$.

For the situation where the detected probabilities are taken to be the
quantum probabilities calculated directly from (\ref{eqn:state}), so that we
are ignoring additional losses and noise which may come from the detection
and measurement process, we have the same result for the weak and strong
inequalities (\ref{eqn:Bellineq}) and (\ref{eqn:weakineq}).

Now to consider detection losses, we notice that the detrimental effect of
the $T$ scaling apparent in (\ref{eqn:loss}) is removed by considering the
weaker inequality, for which the marginal is replaced by the one-sided joint
probability. The quantum predictions for the one-sided probabilities are for
example 
\begin{eqnarray}
P_{++}^{AB}(\theta^{\prime},-)&=&\sum_{m\geq fN}^{N}
\sum_{m^{\prime}=0}^{N}P(m,N-m,m^{\prime},N-m^{\prime})  \nonumber \\
&=&T^{2N}P_{Q}^{A}(m,N-m)  \label{eqn:oneloss}
\end{eqnarray}
which we see from (\ref{eqn:lossamend}) is proportional to $T^{2N}$. Noting
that $P_{Q}(m,N-m)$ is precisely the quantum marginal probability used in
the strong inequality, we see that our predictions then for the violation of
the weak inequality for the state (\ref{eqn:state}) are as shown for the
strong inequality in Figure 3 (meaning that 
the value for $S_{W}$ of equation (\ref{eqn:weakineq}) 
being given by the value of $S$ as shown in Figure 3).

\section{Proposed experiment to detect the violation of the 
multi-particle Bell inequality
using parametric down conversion with and without entangled inputs}

 The prediction by quantum mechanics of the violation of a Bell inequality for
the larger $N$ states (\ref{eqn:state}) has not been 
tested experimentally. For this reason
we investigate how one may achieve related violations of Bell inequalities
using parametric down conversion. Previous work$^{\cite{MB}}$ has shown how
such violations are possible in the regime of low amplification, but
this work was limited to situations of very low detection efficiencies.

We model the parametric down conversion by the Hamiltonian 
\begin{equation}
H=i\hbar g \left(a_1^{\dagger }a_2^{\dagger }+b_1^{\dagger }b_2^{\dagger }
\right) - i\hbar g \left(a_1a_2+b_1b_2 \right)
\end{equation}
Here we consider two parametric processes to make a four mode
interaction$^{\cite{4parexppro}}$, as may be achieved using two 
parametric amplifiers with Hamiltonians 
$H=i\hbar g (a_1^{\dagger }a_2^{\dagger } - 
 a_1a_2)
$ and $H=i\hbar g (b_1^{\dagger }b_2^{\dagger }
 -  b_1b_2)$. The two outputs $a_1, b_1$ are input 
to the polariser $\theta$ at $A$, while the two outputs $a_2, b_2$ are input to 
the polariser $\phi$ at $B$. 
The time-dependent solution for the parametric process with vacuum inputs is 
\begin{equation}
|\varphi \rangle =\sum_{N=0}^{\infty }c_{N}|\varphi _{N}\rangle
\label{eqn:paramp} 
\end{equation}
where $c_{N}=\sqrt{(N+1)}\Gamma ^{N}/\widetilde{C}^{2}$ \ where $\widetilde{C%
}\equiv \cosh r$, $\widetilde{S}\equiv \sinh r$, $\Gamma \equiv \widetilde{S}%
\widetilde{/C}$ and ``gain'': $r=gt$. The probability that a total of $N$
photons are detected at each location $A$ and $B$ is then $P(n)=|c_{N}|^{2}$
as plotted in Figure 5.

\begin{figure}
  \includegraphics[scale=.6]{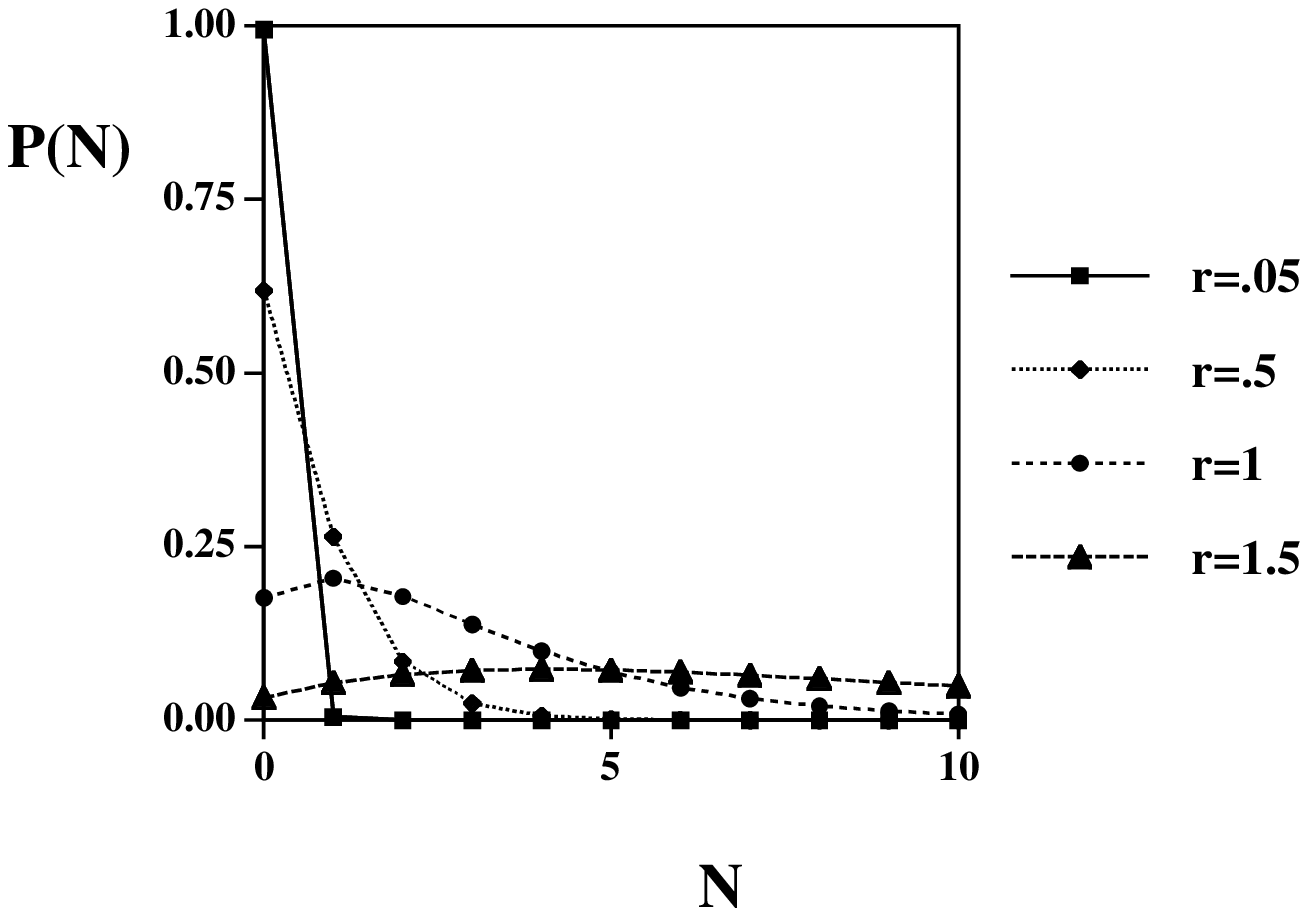}\\
\caption{Plot of $P(N)=|c_{N}|^{2}$ the probability that a total of $N$
photons will be detected at each polariser location.}
\label{fig5}
\end{figure}

Of interest to us is the parametric output with the following
polarization-entangled state as input. 
\begin{equation}
|\varphi \rangle _{in}=\frac{1}{\sqrt{2}}\left( |1\rangle _{a_{1}}|1\rangle
_{a_{2}}|0\rangle _{b_{1}}|0\rangle _{b_{2}}+|0\rangle _{a_{1}}|0\rangle
_{a_{2}}|1\rangle _{b_{1}}|1\rangle _{b_{2}}\right)
\end{equation}
This represents an example of the quantum injected optical parametric
amplifier (QIOPA) realized experimentally by De Martini et al$^{\cite{deM}%
} $. The active NL medium realizing the interaction (18) was a 2 mm BBO
(beta-barium-borate) nonlinear crystal slab excited by a pulsed optical UV
beam with wavelength $\lambda _{p}=345nm$. The duration of each UV
excitation pulses was $150f\sec $ and the average UV power was $0.3W$. The
UV\ beam was SHG\ generated by a mode-locked femtosecond Ti:Sa Laser
(Coherent MIRA) optionally amplified by a high power Ti:Sa Regenerative
Amplifier (Coherent REGA9000). The pulse repetition rate was $76.10^{6}Hz$
and $3.10^{5}Hz$ respectively in absence and in presence of the regenerative
amplification. The maximum OPA ``gain'' obtained by the apparatus was: $%
r\approx 0.3$\ and\ $r\approx 5.1$ respectively in absence and in presence
of \ the laser amplification. These figures lead respectively to the
following values of the parameters: $\widetilde{C}=1.04$, $\Gamma =0.29$ and 
$\widetilde{C}=82$, $\Gamma \approx 1$. The typical quantum efficiency of the
detectors was in the range: $\eta ^{2}\approx 0.4-0.6$. The final output
state generated by this apparatus is expressed by the multi-particle
entangled state (\ref{eqn:paramp}) but where $c_{N}=[\sqrt{(N+1)}\Gamma ^{N}/%
\widetilde{C}^{2}]\times \lbrack (N-2\widetilde{S}^{2})/(\sqrt{2}\Gamma 
\widetilde{C}^{2})]$. The probability of an $n$ photon output at each
location $A$ and $B$ is then given by $P(n)=|c_{n}|^{2}$ as is plotted in
Figure 6, for various $r$.

\begin{figure}
  \includegraphics[scale=.6]{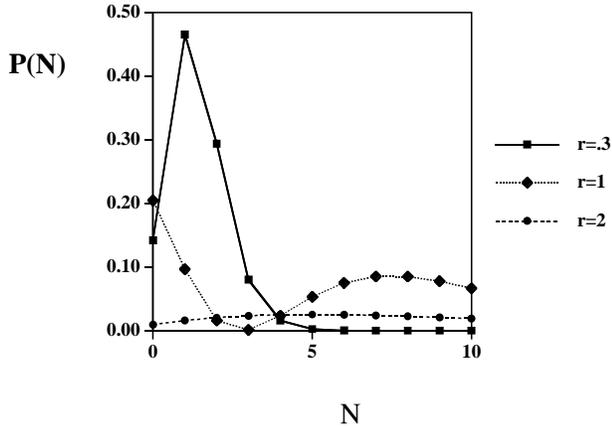}\\
\caption{Plot of $P(N)=|c_{N}|^{2}$ the probability that a total of $N$
photons will be detected at each polariser location, for the 
entangled state input.}
\label{fig6}
\end{figure}

There are a number of approaches one can use to detect the quantum violation
of the Bell inequalities. The particular method preferred will depend on the
interaction strength $r$ and the degree of detection efficiency $\eta$.

\begin{figure}
  \includegraphics[scale=.45]{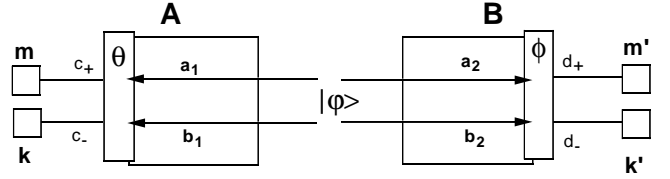}\\
\caption{ Arrangement to test the Bell inequality for the parametric output.
Our outcome is $+1$ at $A$ if $m=N$
and $m+k=N$; and $+1$ at $B$ if $m^{\prime}=N$ and $m^{\prime}+k^{\prime}=N$%
. We measure $P_{++}^{AB}(\protect\theta,\protect\phi)$ and, if 
testing the no loop-hole Bell inequality (5), the marginal
probabilities $P_{+}^{A}(\protect\theta)$ and $P_{+}^{B}(\protect\phi)$. 
If testing the weaker Bell inequality (16), measurement is made of 
one-sided joint probabilities  $P_{+}^{A}(\protect\theta,-)$ and 
 $P_{+}^{b}(-,\protect\phi)$. A
violation of the no auxiliary-assumptions Bell inequality is possible only
for high $T=\protect\eta^{2}$ (Figure 8). Predicted violations of a weak
inequality (16) shown in Figure 9. }
\label{fig7}
\end{figure}

We propose here first the following experiment making use of the
double-channeled polarisers (Figure 7) to detect the photon numbers of both
orthogonal polarisations. This will allow the selection of a specified spin
state $|\varphi_{N}\rangle$ and the observation of the violation predicted
in Figure 3. Specifically we detect at locations $A$ and $B$ the photon
numbers $c_{+}^{\dagger }c_{+}$, $c_{-}^{\dagger }c_{-}$, $d_{+}^{\dagger
}d_{+}$ and $d_{-}^{\dagger }d_{-}$, where $c_{\pm}$ and $d_{\pm}$ are given
by (\ref{eqn:modes}), and label the results $m$, $k$, $m^{\prime}$, and $%
k^{\prime}$ respectively. We designate the result of the measurement at $A$
to be $+1$ if our measured results $m$ and $k$ satisfy $m=N$ and also $m+k=N$%
. Otherwise our result is $-1$. Similarly we define the result at $B$ to be $%
+1$ if $m^{\prime}=N$ and $m^{\prime}+k^{\prime}=N$. By performing many such
measurements over an ensemble, one can experimentally test the 
``strong'' (no loop-hole) Bell
inequality (\ref{eqn:Bellineq}).

The calculation of $S$ as defined in (\ref{eqn:Bellineq}) 
for the parametric amplifier state proceeds in
straightforward manner. We define in general $P_{Q}(m,k,m^{\prime},k^{%
\prime})$ as the probability of detecting $m,k,m^{\prime},k^{\prime}$
photons upon measurements of $c_{+}^{\dagger }c_{+}$, $c_{-}^{\dagger }c_{-}$%
, $d_{+}^{\dagger }d_{+}$ and $d_{-}^{\dagger }d_{-}$ respectively, in the
absence of loss. For $m+k=m^{\prime}+k^{\prime}=N$, we have 
\begin{eqnarray}
P_{Q}(m,N-m,m^{\prime},N-m^{\prime})= |c_{N}|^{2} |C_{m,m^{\prime}}^{(N)}|^2
\label{eqn:quantpar}
\end{eqnarray}
$|c_{N}|^2$ is defined in (\ref{eqn:paramp}), and $|C_{m,m^{%
\prime}}^{(N)}|^2 $ is the probability that measurement of $c_{+}^{\dagger
}c_{+}$ and $d_{+}^{\dagger }d_{+}$ for the state $|\varphi _{N} \rangle $
gives $m$ and $m^{\prime}$ respectively. Our required probabilities are then
given as follows. 
\begin{eqnarray}
P_{++}^{AB}(\theta,\phi) = P_{Q}(N,0,N,0)=|c_{N}|^{2}|C_{N,N}^{(N)}|^2
\end{eqnarray}
and 
\begin{eqnarray}
P_{+}^{A}(\theta) &=&
\sum_{m^{\prime}=0}^{\infty}P_{Q}(N,0,m^{\prime},N-m^{\prime})\nonumber\\
&=&\sum_{m^{\prime}=0}^{\infty}|c_{N}|^{2}|C_{N,m^{\prime}}^{(N)}|^2
\end{eqnarray}

The detection of $m+k=N$ at $A$ is correlated with $m^{\prime}+k^{\prime}=N$
at $B$. Immediately then it is apparent that the factors $|c_{N}|^{2}$ in
the joint and marginal probabilities in the final form of the Bell parameter $%
S$ for the strong inequality 
(\ref{eqn:Bellineq}) will cancel. The predictions for the violation of 
(\ref{eqn:Bellineq}),
in the absence of loss, are as for the ideal spin state $|\varphi_{N}\rangle$%
. It is important to realize however that the actual probability of
obtaining the event $+1$ is different in the parametric case, this
probability being weighted by $|c_{N}|^{2}$, the probability of detecting $%
m+k=N$, that $N$ photons are incident on each polariser. While the joint
probabilities are small, so is the true marginal, and we have a predicted
violation of the strong Bell-Clauser-Horne inequality (\ref{eqn:Bellineq}),
without auxiliary assumptions.

The probabilities $P_{Q}(m,k,m^{\prime},k^{\prime})$ of 
(\ref{eqn:paramp}) 
depend only on the angle difference $\phi -\theta $. 
We select the angle choice $\phi
-\theta=\theta^{\prime}- \phi=\phi^{\prime}-\theta^{\prime}=\psi$ and $%
\phi^{\prime}-\theta=3\psi$ in line with previous work$^{\cite{Drum,MB}}$
with the states $|\varphi_{N} \rangle $.

Our first objective would be to detect violations of the inequality for
relatively low $N$, $N=2$ say. The choice of $r\sim 1$ gives the maximum
probability of obtaining an event where $m+k=2$, although $r\sim.5$ would
give a reasonable probability. For the optimal choice of angle $\psi$
(Figure 3) the probability of an actual event $+1$ for $N=2$ and $r\sim.5$
is $\sim .01$. For perfect detection efficiency the level of violation is
given by $S=1.181$ as indicated in Figure 3.

We now need to consider the effect of detection inefficiencies. Our measured
probabilities for obtaining $m,k,m^{\prime},k^{\prime}$ at each detector are
given by (\ref{eqn:loss}) where now the quantum probabilities are calculated
from (\ref{eqn:paramp}). We note that with the restriction $%
m+k=m^{\prime}+k^{\prime}=N$, and $m=N$ we get 
\begin{eqnarray}
&\quad&P(N,0,m^{\prime},N-m^{\prime})=T^{2N}\sum_{r,q,s,t=0}^{
\infty}(1-T)^{r+q+s+t}  \nonumber \\
&\quad& C_{r}^{N+r} C_{s}^{m^{\prime}+s} C_{t}^{N-m^{\prime}+t}  
 P_{Q}(N+r,q, \nonumber\\
&\quad&\quad\quad\quad\quad\quad m^{\prime}+s,N-m^{\prime}+t)  \label{eqn:lossparamp}
\end{eqnarray}
where from (\ref{eqn:quantpar}) we have 
\begin{eqnarray}
&\quad&P_{Q}(N+r,q,m^{\prime}+s,N-m^{\prime}+t)=  \nonumber \\
&\quad&\delta (r+q-(s+t)) |c_{N_{0}}|^{2} |C_{N+r,m^{\prime}+s}^{(N_{0}?)}|^2
\end{eqnarray}
where $N_{0}=N+r+q=N+s+t$. We note that for the quantum state (\ref
{eqn:quantpar}) we require for nonzero probabilities $r+q=s+t$. The required
joint probability $P_{++}^{AB}(\theta,\phi)$ becomes $P_{++}^{AB}(\theta,%
\phi)=P(N,0,N,0)$. The marginal probabilities needed for the strong Bell
inequality (\ref{eqn:Bellineq}) becomes for example $P_{+}^{A}(%
\theta)=P^{A}(N,0)$ where 
\begin{eqnarray}
P^{A}(N,0)&=&T^{N}\sum_{r,q=0}^{\infty}(1-T)^{r+q}  
 C_{r}^{N+r} P_{Q}^{A}(N+r,q)  \label{eqn:lossparamarg}
\end{eqnarray}
where 
\begin{eqnarray}
P_{Q}^{A}(N+r,q)=\sum_{m^{\prime}=0}^{\infty} |c_{N_{0}}|^{2}
|C_{N+r,m^{\prime}}^{(N_{0}?)}|^2
\end{eqnarray}
where $N_{0}=N+r+q$.

Figure 8 reveals the effect on the violation of the strong Bell inequality,
for various $r$, and for $N=2$. For the reasons discussed in the previous
section, because the marginal probability scales as $T^{N}$ while the joint
probabilities scale as $T^{2N}$, the violation is lost for small detection
loss.

\begin{figure}
  \includegraphics[scale=.6]{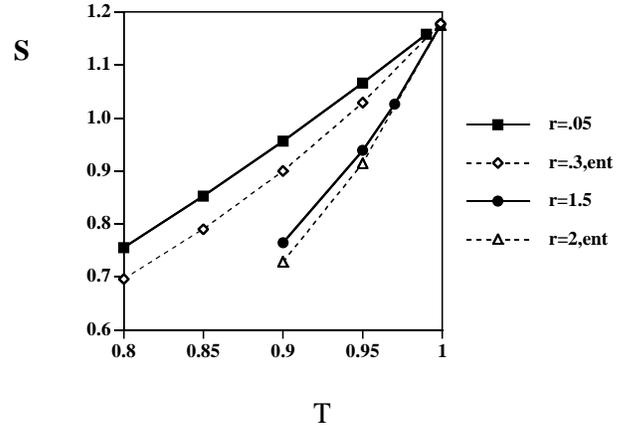}\\
\caption{Effect, for various parametric coupling $r$, of detection
inefficiencies on the violation of the strong Bell inequality (5), for 
the scheme described in Figure 7 with $N=2$. Here $T$ models detector 
losses, $T$ being 
the relative fraction of photons incident on each detector that
are actually detected. The optimal angle $\protect\psi$ for $N=2$ is $\sim
3.41$. The curves labeled ``ent'' represent predictions for the 
entangled input state.}
\label{fig8}
\end{figure}

To propose an experiment achievable with current detector efficiencies, we
consider an appropriate weak Bell inequality. We define as before (Figure 7)
the joint probability $P_{++}^{AB}(\theta,\phi)$ of obtaining $m=N$ and $%
m+k=N$ at $A$, and $m^{\prime}=N$ and $m^{\prime}+k^{\prime}=N$ at $B$. We
define the joint one-sided probability $P_{++}^{AB}(\theta,-)$ of obtaining $%
m=N$ and $m+k=N$, and a total of $m^{\prime}+k^{\prime}=N$ photons at $B$.
The one-sided probability $P_{++}^{AB}(-,\phi)$ is defined similarly. The
auxiliary assumptions are made that for a hidden variable description $%
\lambda$, the probability $p_{+}^{A}(\theta,\lambda)$ of obtaining $m=N$ and 
$m+k=N$, and the probability $p_{+}^{A}(-,\lambda)$ of obtaining $m+k=N$
alone, satisfy 
\begin{eqnarray}
p_{+}^{A}(\theta,\lambda) \leq p_{+}^{A}(-,\lambda).  \label{eqn:aux}
\end{eqnarray}
Also we assume $p_{+}^{A}(-,\lambda)$ is independent of $\theta$. Similar
assumptions are made for $p_{+}^{B}(\phi,\lambda)$ and $p_{+}^{B}(-,\lambda)$%
. With these assumptions the weaker inequality (\ref{eqn:weakineq}) is
derivable. The one-sided probability used in the test of the weak inequality
(\ref{eqn:weakineq}) is given by $P_{++}^{AB}(\theta,-)=P^{AB}(N,0;-)$ where 
\begin{eqnarray}
P^{AB}(N,0;-)&=&\sum_{m^{\prime}=0}^{N}P(N,0,m^{\prime},N-m^{\prime})
\label{eqn:lossparampone}
\end{eqnarray}
With a total of $N$ photons detected at both locations $A$ and $B$, we
ensure all probabilities scale as $T^{2N}$.

The existence of the higher spin states $|\varphi_{M}\rangle$ where $M>N$ in
the parametric output means that detector inefficiencies alter the violation
of even the weak Bell inequality. Figure 9 illustrates the effect of
detection inefficiencies on the violation of the weak inequalities (\ref
{eqn:weakineq}), the effect being more significant for higher $r$ values
where the states $|\varphi_{M}\rangle$, where $M>N$, contribute more
significantly. Smaller $r$ values suffer the disadvantage however that the
probability of an actual event $+1$ becomes small due to the small
probability of $N=2$ photons actually being incident on the polariser. The
sensitivity of the violations to loss is not so great that the experiment
would be impossible for $r\sim .5$.

\begin{figure}
  \includegraphics[scale=.6]{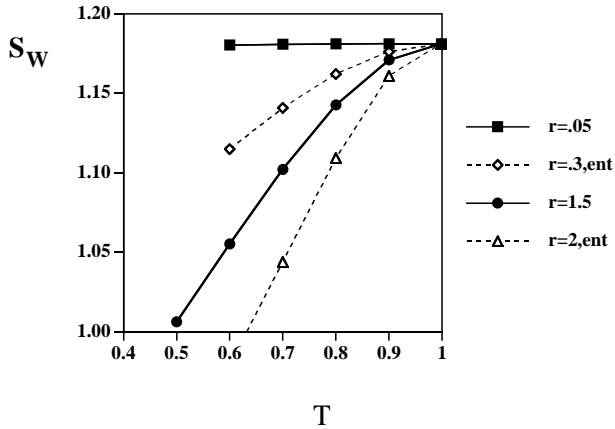}\\
\caption{Effect of detection inefficiencies on the violation of the weak
Bell inequality (16), for the scheme described in figure 7 where $N=2$.
 Here $T$ represents detector losses, $T$ being the relative fraction of incident
photons actually detected. The optimal angle $\protect\psi$ for $N=2$ is $%
\sim 3.41$. The curves labeled ``ent'' are predictions for the entangled 
input state. }
\label{fig9}
\end{figure}

A point to be made concerns the alternative situation of a one-channeled
polariser where only the photon number $m$ and $m^{\prime}$ can be detected.
Here the prediction is different due to the contribution of the $N+1$ spin
state which can contribute an $m=N$ event (with $k=1$) potentially
decreasing the violation of the inequality.

\section{Proposed experiment to detect the violation of the Bell inequality
using high flux parametric down conversion}

As one increases the output intensities of the parametric device, the actual
probability of detecting $N$ photons transmitted through our polariser
decreases, in other words the probability of detecting the event $+1$,
described in the last section, becomes smaller. To combat this we propose in
this section that our outcome be a range of photon number values. Here we
are interested in the regime of high amplification$^{\cite{Smithey}}$ where
the output fluxes of signal and idler are high, and where one can use highly
efficient photodiode detectors.

We now propose the following experiment. We detect at locations $A$ and $B$
the photon numbers $c_{+}^{\dagger }c_{+}$, $c_{-}^{\dagger }c_{-}$, $%
d_{+}^{\dagger }d_{+}$ and $d_{-}^{\dagger }d_{-}$, where $c_{\pm}$ and $%
d_{\pm}$ are given by (3). The mean photon number incident on each polariser
is $xm=<c_{+}^{\dagger }c_{+}>+<c_{-}^{\dagger }c_{-}> =<d_{+}^{\dagger
}d_{+}>+<d_{-}^{\dagger }d_{-}>$ where $xm=2sinh^{2}(r)$. We denote the
result for $c_{+}^{\dagger }c_{+}$ and $c_{-}^{\dagger }c_{-}$ at $A$ by $m$
and $k$ respectively, and the results of $d_{+}^{\dagger }d_{+}$ and $%
d_{-}^{\dagger }d_{-}$ at $B$ by $m^{\prime}$ and $k^{\prime}$ respectively
(Figure 7). We define $XM$ to be the integer nearest in value to the mean $%
xm $. We designate the result of the measurement at $A$ to be $+1$ if our
measured results $m$ and $k$ satisfy $m \geq XM$ and also $XM\leq m+k \leq
XM+\Delta$. Otherwise our result is $-1$. Similarly we define the result at $%
B$ to be $+1$ if $m^{\prime}\geq XM$ and $XM \leq m^{\prime}+k^{\prime}\leq
XM+\Delta$.

By performing many such measurements over an ensemble, one can
experimentally determine the following: $P_{++}^{AB}\left( \theta ,\phi
\right) $ the probability of obtaining $+1$ at $A$ and $+1$ at $B$ upon
simultaneous measurement with $\theta$ at $A$ and $\phi $ at $B$; $%
P_{+}^A\left( \theta \right)$ the marginal probability for obtaining the
result $+1$ upon measurement with $\theta$ at $A$; and $P_{+}^B\left( \phi
\right)$ the marginal probability of obtaining the result $+1$ upon
measurement with $\phi$ at $B$.

Local hidden variables will predict as discussed in section 2 the strong
Bell inequality (\ref{eqn:Bellineq}). We define $P(m,k,m^{\prime},k^{%
\prime}) $ as the probability of detecting $m,k,m^{\prime}$and $k^{\prime}$
photons for measurements of $c_{+}^{\dagger }c_{+}$, $c_{-}^{\dagger }c_{-}$%
, $d_{+}^{\dagger }d_{+}$ and $d_{-}^{\dagger }d_{-}$ respectively. The
probability of results $m$ and $k$ upon measurement of $c_{+}^{\dagger
}c_{+} $and $c_{-}^{\dagger }c_{-}$ is defined as $P^{A}(m,k)$. We have in
the absence of loss, where $m+k=m^{\prime}+k^{\prime}=N$ is ensured, 
\begin{eqnarray}
P(m,N-m,m^{\prime},N-m^{\prime}) &=& |c_{N}|^{2} |C_{m,m^{\prime}}^{(N)}|^2 
\nonumber \\
P^{A}(m,N-m)&=&\sum_{m^{\prime}=0}^{N}|c_{N}|^{2} |C_{m,m^{\prime}}^{(N)}|^2
\end{eqnarray}
where all other probabilities are zero. Here $|c_{N}|^2$ is defined in (\ref
{eqn:paramp}), and $|C_{m,m^{\prime}}^{(N)}|^2$ is the probability that
measurement of $c_{+}^{\dagger }c_{+}$ and $d_{+}^{\dagger }d_{+}$ for the
state $|\varphi _{N} \rangle $ gives $m$ and $m^{\prime}$ respectively, with
no loss. The probability $P(m,m^{\prime})$ of getting $m$ and $m^{\prime}$
for $c_{+}^{\dagger }c_{+}$ and $d_{+}^{\dagger }d_{+}$ respectively, while
the total $m+k$ is restricted to $XM\leq m+k \leq XM+\Delta$, and $%
m+k=m^{\prime}+k^{\prime}$ is restricted to $XM \leq
m^{\prime}+k^{\prime}\leq XM+\Delta$, is given generally as 
\begin{eqnarray}
P^{AB}(m,m^{\prime}) = \sum_{k=XM-m}^{XM+\Delta-m}
\sum_{k^{\prime}=XM-m^{\prime}}^{XM+\Delta-m^{\prime}}P(m,k,m^{\prime},k^{%
\prime})  \label{eqn:rangeprob}
\end{eqnarray}
The corresponding marginal probability is 
\begin{eqnarray}
P^{A}(m) = \sum_{k=XM-m}^{XM+\Delta-m} P^{A}(m,k)
\end{eqnarray}
Our required probabilities are then given as follows 
\begin{eqnarray}
P_{++}^{AB}(\theta,\phi) = \sum_{m,m^{\prime}=XM}^{XM+\Delta}
P^{AB}(m,m^{\prime})
\end{eqnarray}
and for the marginal 
\begin{eqnarray}
P_{+}^{A}(\theta) = \sum_{m=XM}^{XM+\Delta} P^{A}(m)
\end{eqnarray}
For the purpose of a weaker Bell inequality we also define a one-sided
probability 
\begin{eqnarray}
P_{++}^{AB}(\theta,-) =
\sum_{m=XM}^{XM+\Delta}\sum_{m^{\prime}=0}^{XM+\Delta} P^{AB}(m,m^{\prime})
\end{eqnarray}

The probabilities $P(m,k,m^{\prime},k^{\prime})$ depend only on the angle
difference $\phi -\theta $. We select the angle choice $\phi
-\theta=\theta^{\prime}- \phi=\phi^{\prime}-\theta^{\prime}=\psi$ and $%
\phi^{\prime}-\theta=3\psi$ in line with previous work$^{\cite{Bell,Drum}}$
with the states $|\varphi_{N} \rangle $.

Results for $S$, optimizing $\psi$ to give maximum $S$, are presented in the
Figure 9. With the choice $\Delta=0$, we will get only one of the $%
|\varphi_{N}\rangle $ contributing. The results for $S$ will be identical$^{
\cite{Drum}}$ to that obtained for the $|\varphi_{XM}\rangle$ state, where a
clear violation of the Bell inequality (\ref{eqn:Bellineq}) is obtained even
for very large $N=XM$. The difficulty with such a situation however is that
in the regime of higher $r$ where greater signal intensities are generated
the probability that the total number $m+k$ of photons $c_{+}^{\dagger
}c_{+} + c_{-}^{\dagger }c_{-}$ is just this fixed number is very small,
making the probability of our $+1$ outcome tiny. We are more interested in
situations where the intensity on the detectors is large but also where the
probability that $XM\leq m+k \leq XM+\Delta$ is significant. This is
achieved by increasing the range $\Delta$. Violations of the Bell inequality
are still possible ($S\geq1$) but the degree of violation is reduced, the
limiting value for large $\Delta$ approaching $1$ as $XM$ increases.

\begin{figure}
  \includegraphics[scale=.6]{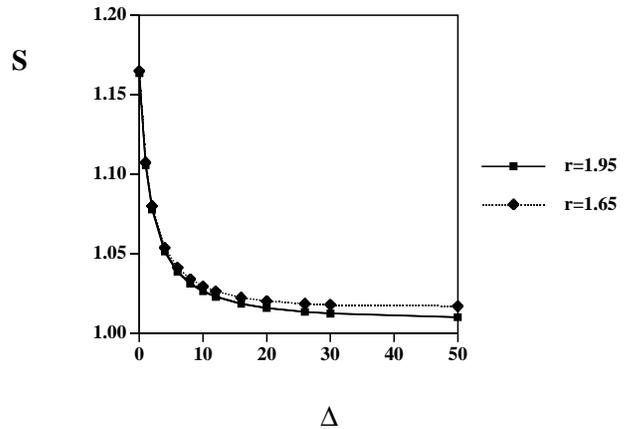}\\
\caption{Plot of violation of the strong Bell inequality where we designate
the result of the measurement at $A$ to be $+1$ if our measured results $m$
and $k$ satisfy $m \geq XM$ and also $XM-\Delta \leq m+k \leq XM+\Delta$.
Otherwise our result is $-1$. Similarly we define the result at $B$ to be $%
+1 $ if $m^{\prime}\geq XM$ and $XM \leq m^{\prime}+k^{\prime}\leq XM+\Delta$%
. For $r=1.65$ we have $XM=13$ and for $r=1.95$, $XM=24$. }
\label{fig10}
\end{figure}
\begin{figure}
  \includegraphics[scale=.6]{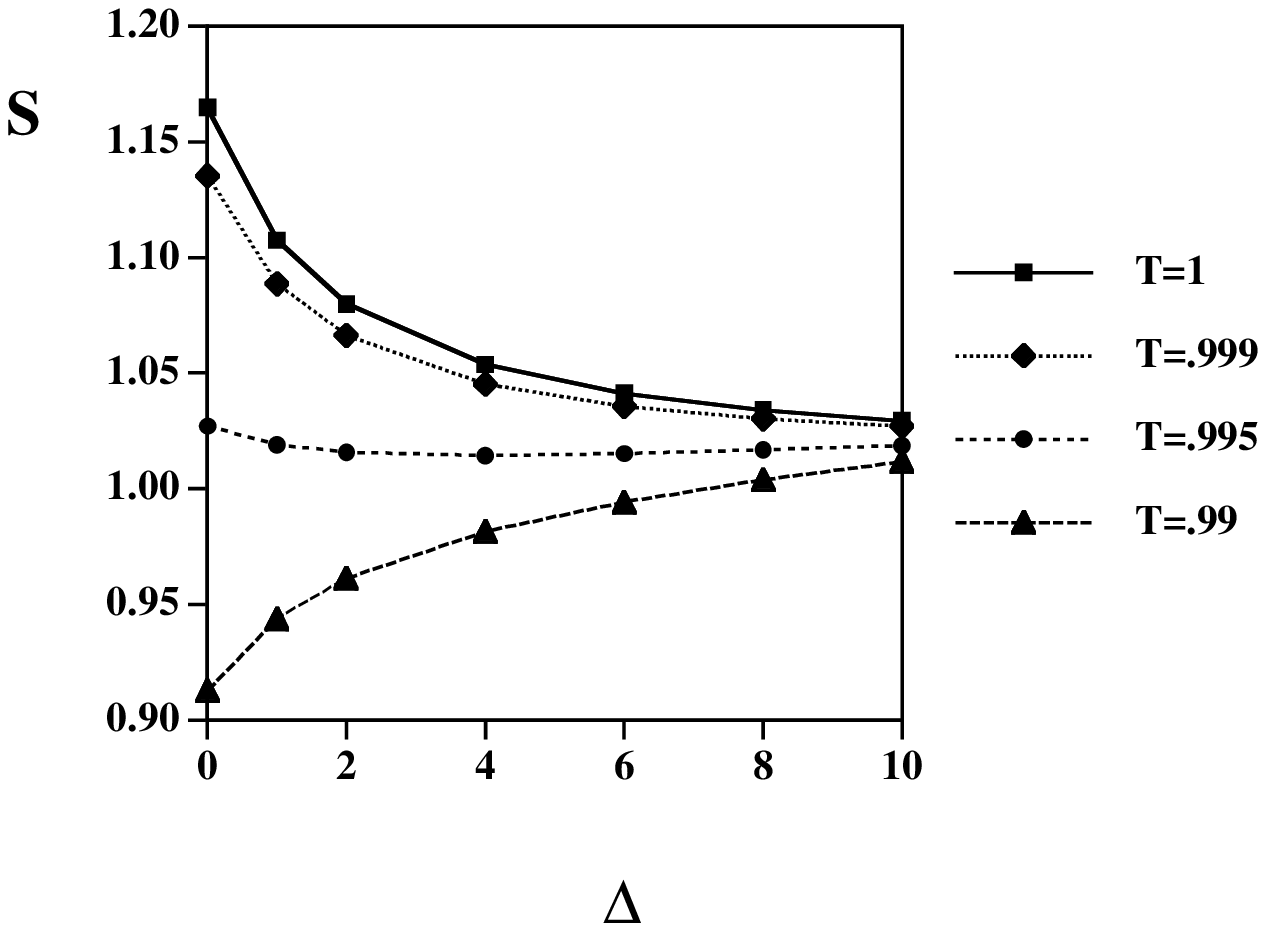}\\
\caption{Plot of the effect of detection losses on the violation of the
no loop-hole Bell inequality test (5) as explained in figure 10 above. Here $r=1.65$
we have $XM=13$. }
\label{fig11}
\end{figure}
\begin{figure}
  \includegraphics[scale=.6]{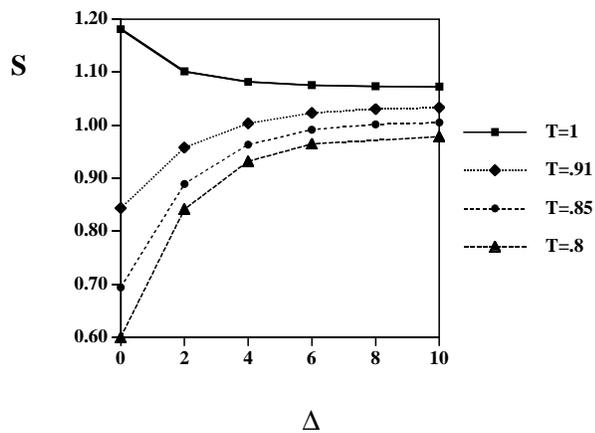}\\
\caption{Plot of the effect of detection losses on the violation of the
no loop-hole Bell inequality test (5) as explained in figure 10 above. Here $r=0.9$ we
have $XM=2$. }
\label{fig12}
\end{figure}

The sensitivity to loss can be evaluated by calculating in (\ref
{eqn:rangeprob}) (and in the equations for the marginal probabilities such
as $P^{A}(m)$) the measured probabilities $P(m,k,m^{\prime},k^{\prime})$ and 
$P^{A}(m,k)$ as given by (\ref{eqn:loss}) and (\ref{eqn:margloss}). The
effect on the violation of the no loop-hole Bell inequality 
(\ref{eqn:Bellineq}) 
is given in Figures 11 and 12.
Sensitivity is strong for low $\Delta$ but decreases as the range $\Delta$
increases. This provides a potential opportunity to test a strong
no-auxiliary multi-particle Bell inequality for lower detector
inefficiencies than indicated by the $\Delta=0$ regime discussed in the
previous section.

\section{Conclusions}
We have presented a proposal
 to test the predictions of quantum mechanics against those 
of local hidden variable theories for multi-particle entangled 
states generated using parametric down conversion, where measurement 
is made on systems of more than one particle. 
A calculation is  
given of the detector efficiencies required to test directly the ``no 
loop-hole''multi-particle Bell 
inequality. In view of the limitation of current 
detector efficiencies, it is necessary to consider initially 
tests of a ``weaker'' Bell 
inequality derived with additional auxiliary assumptions, and to 
therefore extend previous such derivations to the multi-particle situation we 
consider here.  

\bigskip

\bigskip

\section{\protect\bigskip Acknowledgments}

F.D.M. acknowledges the Italian Ministero dell'Universita' e della Ricerca
Scientifica e Tecnologica (MURST) and the FET European Network
IST-2000-29681(ATESIT) on ``Quantum Information\bigskip '', for funding.

\end{document}